\documentclass[twocolumn,showpacs,preprintnumbers,amsmath,amssymb]{revtex4}

\usepackage[T2A]{fontenc}
\usepackage{graphicx}
\usepackage{dcolumn}
\usepackage{bm}

\begin{document}

\title{Dissipative light field as a way to create strongly localized structures for atom
lithography}

\author{O.N. Prudnikov, A.V. Taichenachev}
\affiliation{Novosibirsk State University, Pirogova 2, Novosibirsk 630090, Russia}
\author{A.M. Tumaikin, V.I. Yudin}
\affiliation{Institute of Laser Physics SB RAS, Lavrentyeva 13/3, Novosibirsk 630090,
Russia}

\begin{abstract}
Generally, the conditions for deep sub-Doppler laser cooling do not match the conditions
for the strong atomic localization that takes a place in deeper optical potential and, in
consequence, leads to larger temperature. Moreover, for a given detuning in a deep
optical potential the secular approximation which is usually used for quantum
description of laser cooling becomes no more valid. Here we perform an analysis of atomic
localization in optical potential based on a full quantum approach for atomic density
matrix. We also show that the laser cooling in a deep far-off detuned optical potential,
created by a light field with a polarization gradient, can be used as an alternative
method for forming high contrast spatially localized structures of atoms for the purposes
of atom lithography and atomic nanofabrication. Finally, we perform an analysis of the
possible limits for the width and the contrast of localized atomic structures that can in
principle be reached by this type of the light mask.
\end{abstract}

\pacs{32.80.Pj, 42.50.Vk}

\maketitle

\section{Introduction}
Laser cooling and manipulation of neutral atoms is one of the priority field of atom
optics. In the recent time the major developments and success have been obtained in atom
lithography and direct deposition of atoms utilizing light fields as an immaterial
optical mask for atomic beam \cite{Meschede03,Oberthaler03}. In most nanofabrication
experiments, atomic structures are realized by far off detuned periodical conservative
potential created by intense laser fields acting as an array of immaterial light lenses
for the atoms. The influence of the spontaneous emission on the focusing is considered
to be negligible because of the large light detuning and short interaction times. In
essence, the atom trajectory affected by conservative dipole force without any loses (or
dissipation) of energy in the atomic beam. In this case the atomic beam focusing has a
classical analogy and can be described with methods developed for particle optics
\cite{McClelland91}. As in normal optics, the feature size is limited by a combination
of chromatic aberration caused by the broad longitudinal and transverse velocity
distribution of an atomic beam. Therefore an additional laser cooling field is required
to prepare a well collimated and transversely cooled atomic beam to minimize deleterious
effects. Additionally, because of spherical aberration some atoms do not focus well and
contribute to pedestal background. These factors are dominant and do not allow one to
reach the theoretically predicted diffraction limit for atom optics determined by de
Broglie wavelength of atoms, only a few picometres. Therefore the new alternative
methods for atom lithography are intensively investigated.

Recently, the idea of combining the traditional focusing method with the well known
concept of laser cooling was suggested for a blue detuned intensive light field
\cite{Stutzle03,Pru04} and was mentioned earlier in \cite{kaz}. Here the intense light
field, commonly used for focusing, was used to create a deep optical potential.
Additional dissipative light force provides additional cooling of the atoms to the
minimum of optical potential at blue detuning. The characteristic time when the
dissipation processes take effect is a few inverse recoil frequencies $\omega_{R}^{-1}$
(where $\hbar \, \omega_{R}=\hbar^2 k^2/2M$ is recoil energy, gained by an atom with
mass $M$ at rest after emission of a photon with momentum $\hbar k$). This time is
several tens of microseconds for the number of elements with closed dipole optical
transitions suitable for laser cooling. Thus for commonly used atomic beams with thermal
longitudinal velocity it might be difficult to realize this type of dissiaptive optical
mask experimentally due to power limitations of laser system used.

In the following we consider the alternative regime of dissipative optical mask, created
by red detuned low intensity light field with nonuniform polarization. It is well known
that low intensity light field with polarization gradients can be used for sub-Doppler
laser cooling of neutral atoms. This mechanism of laser cooling is well understood and
has been thoroughly studied by a number of authors
\cite{Wineland79,Dal89,Juha92,Berman93, Castin94} especially with respect to the
temperature of laser cooling, the atomic momentum distribution, and the localization
\cite{Gatzke97,Marksteiner95} that can be measured by spectroscopy methods
\cite{Jessen92,Raithel97}. Due to the extremely complex master equation for the quantum
description of atomic motion in light fields the secular approximation
\cite{Dal91,Dal93,Berman93,Castin94,Deutsch97} was initially suggested is valid in the
limit \cite{Dal91}
\begin{equation}\label{sec}
  \sqrt{U_0/\hbar \omega_R} \ll |\delta|/\gamma \, .
\end{equation}
This limit assumes the energy separation between different energy bands of atoms in
optical potential are much greater then their width due to optical pumping and tunneling
effects. Here the light-shift well depth $U_0$ define the optical potential depth.
$\delta =\omega-\omega_0$ is the detuning between the laser $\omega$ and atomic
transition $\omega_0$ frequencies, and $\gamma$ is the radiative decay rate. This
approximation is valid, for a given potential depth, in the limit of large detuning. On
the contrary, it might failed in a deep potential at given detuning. More over, even if
the secular approximation is well fulfilled for the lowest vibrational levels it might
break down for upper ones where the separation between energy bands becomes smaller due
to potential anharmonicity effects, and especially for above-barrier motioned atoms.

In the present work we investigate the applicability of laser cooling in a deep optical
potential created by light field with nonuniform polarization for generating spatially
localized atom structures with high contrast for atom lithography. We consider the
conditions far from the situation of extremely low sub-Doppler cooling cases. Thus, to
describe the localization of atoms more correctly in here we do not restrict our
consideration by secular approximation. Rather we perform a full quantum numerical
analysis of generalized optical Bloch equation for atomic density matrix elements. In
particular, we consider the light field parameters beyond the secular approximation
limit. Finally we analyze the structure width and contrast of localized atoms, important
parameters for technological applications.

\section{Master equations}
Let us consider one-dimensional (along the z axis) motion of atoms
with total angular momenta $j_g$ in the ground state and $j_e$ in
the excited state in a field of two oppositely propagating waves
with the same frequency and intensity
\begin{eqnarray}\label{field}
  {\bf E}(z,t) &=& E_{0}\left({\bf e}_1 \,e^{ikz} +
  {\bf e}_2\, e^{-ikz}\right)e^{-i\omega t} + c.c. \nonumber \\
&& {\bf e}_{n} = \sum_{q =0,\pm 1} e^{q}_{n} {\bf e}_q \,, \,\,\,
n=1,2
\end{eqnarray}
Here $E_0$ is the amplitude of each of the oppositely propagating
waves. The unit vectors ${\bf e}_1$ and ${\bf e}_2$ determine
their polarizations with components $e^{q}_{n}$ in cyclic basis
$\{ {\bf e}_0 = {\bf e}_z, {\bf e}_{\pm 1} = \mp ({\bf e}_x \pm
i\, \bf{e}_y)/\sqrt{2} \}$.

In this work we restrict our consideretion by the weak-field
limit, i.e. with small saturation parameter
\begin{equation}\label{sat}
  S = \frac{\Omega^2}{\delta^2+\gamma^2/4} \,.
\end{equation}
Here $\Omega = -E_0d/\hbar$ is the single-beam Rabi frequency
characterize the coupling between the atomic dipole $d$ and light
field.

 In the weak-field limit, the atomic exited-state can be
adiabatically eliminated, and atom motion is describing by
reduced equation for the ground-state density matrix elements
\cite{Berman93,Castin94}:
\begin{equation}\label{equation}
 \frac{d}{d t} {\hat \rho} = -\frac{i}{\hbar}\left[{\hat H}, {\hat \rho}
 \right]+{\hat \Gamma}\{{\hat \rho}\}
\end{equation}
where the Hamiltonian $\hat{H}$ is given by
\begin{equation}\label{ham}
  {\hat H} = \frac{{\hat p}^2}{2M} + \hbar\,\delta S \;{\hat V}^{{\dagger}}{\hat
  V}\,.
\end{equation}
The last term in (\ref{ham}) describes atoms interaction with the
light field in resonance approximation, where
\begin{eqnarray}\label{Vmat}
 \hat{V} &=& \hat{V}_1\, e^{ikz}+\hat{V}_2\, e^{-ikz}  \nonumber \\
 &=& \sum_{q} \hat{T}_q \,e^{q}_1 \;e^{ikz} +
\sum_{q} \hat{T}_q\,  e^{q}_2\; e^{-ikz}
 \, ,
\end{eqnarray}
and operator  $\hat{T}_{q}$ is written through the Clebsch-Gordan
coefficients:
\begin{equation}\label{T}
\hat{T}_{q} = \sum_{\mu_e, \mu_g} C^{je, \mu_e}_{1,q;\, j_q,
  \mu_g} |j_e, \mu_e \rangle \langle j_g, \mu_g|
\end{equation}
written in the basis of sublevel wave functions for exited $|j_e, \mu_e \rangle$ and the ground
$|j_g, \mu_g \rangle$ atomic states.

In addition, the relaxation part of kinetic equation for atomic
density matrix (\ref{equation}) has the following form
\begin{eqnarray}\label{relaks}
&&\hat{\Gamma}\{\hat{\rho}\} = -\frac{\gamma S}{2}
\left\{ \hat{V}^{\dagger} \hat{V}, \hat{\rho} \right\}  \\
&+&\gamma S  \sum_{q=0,\pm 1 }\int_{-1}^{1} {\hat T}_q^{\dagger}
e^{-i k s \hat{z}} \, \hat{V} \hat{\rho}\,\hat{V}^{\dagger} e^{-i
k s \hat{z}}{\hat T}_q K_q(s)\, ds \nonumber
\end{eqnarray}
where $\{\hat{a},\hat{c}\} = \hat{a}\hat{c}+\hat{c}\hat{a}$
standard anticommutator definition and ${\hat z}$ is position
operator. This term describes redistribution of the atom on the
ground state energy sublevels with taking into account the recoil
effects in spontaneous photon emission. The functions $K_{\pm
1}(s) = (1+s^2)3/8$ and $K_{0}(s) = (1-s^2)3/4$ is determined by
the probability of emission of a photon with polarization $q=\pm 1
,0$ into direction $s = \cos(\theta)$ (relative to the $z$ axis).

\section{Equilibrium atomic density matrix}
There are number of approaches developed for calculation of the evolution of the atomic
density matrix. The quantum problem is more difficult because it incorporate evolution
of numbers of internal and external components of the density matrix. The majority of
works are based on secular approximation for density matrix elements \cite{Berman93,
Castin94,Deutsch97}. It consists in the following: First, the eigenstates and
eigenenergies of the Hamiltonian $\hat{H}$ should be found. Then we can consider only the
evolution of diagonal elements of atomic density matrix in this eigenstatas basis. It
can be written in a form of balance equation with the rates characterizing the relaxation
part of the master equation. The secular approximation for the general master equation
(\ref{equation}) is valid for the lowest vibrational energy levels when the the energy
separation between different bands are much greater than the effective width due to
optical width and tunneling effects. This condition implies very large detuning.
However, the density of energy states increases for the upper energy levels and the
energy difference between adjacent states could be very small \cite{Castin94}.  This
circumstances therefore make very hard to use the secular approximation for describing
of hot and nonlocalized atomic fraction.

In order to take into account effects of localization in optical potential and modulation
depth of the spatial distribution of atoms more correctly we utilize a different approach
for the quantum calculation of the master equation (\ref{equation}).

In the Wigner representation for atomic density matrix
$\hat{\rho}(x,p)$ the general master equation (\ref{equation})
takes the following form:
\begin{equation}\label{wigner}
  \frac{d}{dt} \hat{\rho} =
  -i \delta \, S \left[\hat{V}^{\dagger} \hat{V}, \hat{\rho}\right]
  - \frac{\gamma \,S}{2} \left\{\hat{V}^{\dagger} \hat{V}, \hat{\rho}\right\} +\hat{\gamma}\{\hat{\rho}\}
\end{equation}
where commutator of density matrix with kinetic part of Hamiltonian (\ref{ham}) reduces
to partial deviation on $z$
\begin{equation}\label{dt}
  \frac{d}{dt} \hat{\rho}(z,p) \equiv \left( \frac{\partial}{\partial t}
  + \frac{p}{M} \frac{\partial}{\partial z} \right) \hat{\rho}(z,p)
\end{equation}
The field part of Hamiltonian (\ref{ham}) has only the zeroth and
the second spatial harmonics:
\begin{equation}\label{Wharm}
\hat{V}^{\dagger} \hat{V} = \hat{W}_0 + \hat{W}_{+} e^{i2kz} +
\hat{W}_{-} e^{-i2kz}
\end{equation}
thus the commutator and anticommutator in the right hand side of equation (\ref{wigner})
could be written as:
\begin{eqnarray}\label{Wpart}
\hat{V}^{\dagger} \hat{V} \, \hat{\rho} &\mp& \hat{\rho}\,\,
\hat{V}^{\dagger}
 \hat{V}= \hat{W}_0 \,\hat{\rho}(z,p) \mp \hat{\rho}(z,p)
 \hat{W}_0 \\
 &+& \left(\hat{W}_{-} \, \hat{\rho}(z,p+\hbar k) \mp
 \hat{\rho}(z,p-\hbar k)\hat{W}_{-}
 \right)e^{-i2kz} \nonumber \\
 &+& \left(\hat{W}_{+} \, \hat{\rho}(z,p-\hbar k) \mp
 \hat{\rho}(z,p+\hbar k)\hat{W}_{+}
 \right)e^{i2kz} \nonumber \, .
\end{eqnarray}
The last term describing relaxation due to spontaneous emission of photons has
well-known form in Wigner representation:
\begin{eqnarray}\label{Wrelax}
  \hat{\gamma}\{\hat{\rho}(z,p)\} = \gamma S \sum_{q=0,\pm 1 }\int_{-\hbar k}^{\hbar k} dp'/\hbar k\; \;
   K_q(p'/\hbar k) \;\; {\hat T}_q^{\dagger}
\nonumber \\
\times \left [
\hat{V}_1\, \hat{\rho}(z,p+p') \; \hat{V}_2^{\dagger}e^{i2kz} +
\hat{V}_2\, \hat{\rho}(z,p+p') \; \hat{V}_1^{\dagger} e^{-i2kz}
\right. \nonumber \\
+ \left.
\hat{V}_1\, \hat{\rho}(z,p+p'-\hbar k) \;
\hat{V}_1^{\dagger} +
\hat{V}_2\, \hat{\rho}(z,p+p'+\hbar k) \; \hat{V}_2^{\dagger}
\right]
{\hat T}_q \, . \nonumber
\end{eqnarray}
Equation (\ref{wigner}) admits solution that is periodic in the
position variable. We further introduce a Fourier-series
expantion for atomic density matrix in spatial coordinates
\begin{equation}\label{fourier}
  \hat{\rho}(z,p) = \sum_n \hat{\rho}^{(n)}(p) \; e^{i\,2n\,kz}
\end{equation}
and rewrite the master equation for discrete Fourier components
of density matrix $\hat{\rho}^{(n)}$:
\begin{eqnarray}\label{Lharm}
\left(\frac{\partial}{\partial t} \right.&+& \left.
2ni\,\frac{p}{M}\right)\hat{\rho}^{(n)} =  \\
&& {\cal L}_{0} \left\{ \hat{\rho}^{(n)} \right\}
  +{\cal L}_{+} \left\{ \hat{\rho}^{(n-1)} \right\}
  +{\cal L}_{-} \left\{ \hat{\rho}^{(n+1)} \right\} \, . \nonumber
\end{eqnarray}

\begin{figure}[pt]
\begin{center}
\includegraphics[width= 3.0 in]{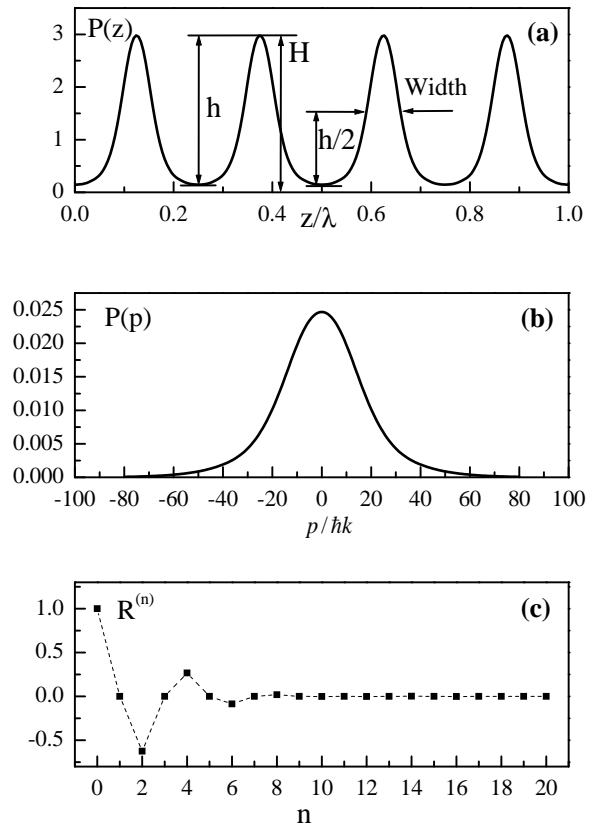}
\end{center}
\caption{\em Steady-state spatial (a) and momentum (b) distributions, and the total population
of spatial harmonics of atomic ground-state density matrix (c) for atoms with $j_g =
1 \to j_e = 2$ optical transition and mass as for $Cr$ atoms.
The light field detuning $\delta = -40 \gamma$ and saturation parameter $S = 0.5$.} \label{fig1}
\end{figure}

 In steady-state problem ($\partial/\partial t \; \hat{\rho} = 0$) such
a recursion may often be solved by the method of continued fraction. This approach is
used for solution of the optical-Bloch equations in different spectroscopy tasks as well
as for the calculation of the force on atom in the light field (see for example
\cite{Risk,Tan}). The major distinction here is that these equations for density matrix
contain the recoil effects that makes them more complicated for calculation.
Additionally we note, the similar approach was described in \cite{Juha91} where the
authors analyzed the laser cooling (velocity distribution) of two-level atom in the
recoil limit and thus restrict their consideration only by the zeroth spatial harmonics
for the ground state density matrix. In our case the number of the spatial harmonics
depends on the light field and atomic parameters. Typically we use less than $30$
harmonics that is enough to obtain the spatial solution for equilibrium atomic density
matrix in considered range of light field parameters.

The spatial and momentum steady-state distribution of atoms with $j_g =
1 \to j_e = 2$ optical transition for $\delta = -40 \gamma$, $S = 0.5$ and chromium mass
are shown in Fig.\ref{fig1}(a) and (b). The Fig.\ref{fig1}(c) represents
the total population of spatial harmonics of atomic ground-state
density matrix integrated over momentum space $R^{(n)} = \int Tr\{\hat{\rho}^{(n)}(p)\}\, dp$.
The zeroth harmonic is equal to $1$ that is the normalization condition. As
it seen here the population of higher harmonics are rapidly decrease with number $n$.

\section{Results}
In this section we turn our attention to the steady-state spatial distribution of the
atoms in the optical potential created by the light field with $lin \perp lin$
configuration. We choose this configuration as a brightest example of the light field
with nonuniform polarization. Only light field ellipticity varies in position space
while the other parameters (intensity, phase, orientation of polarization vector) stay
unchanged. More over, the optical potential created by this field configuration has a
period of $\lambda/4$ that makes it very attractive for deposition of atomic structures
with high spatial periodicity.

\begin{table}[tbp]
\begin{center}
\begin{tabular}{ccccccccc}
\hline \hline \\
Element && cooling &&$\tilde{M}$ &&  $\lambda$ && $I_S$\\
         &&transition &&         &&  (nm) && $(mW/cm^2)$\\ \\
\hline \\
$^{7}$Li  &&  $2^2S_{1/2} \to 2^2P_{3/2}$   &&   46  &&  671   &&  2.56 \\
$^{23}$Na &&  $3^2S_{1/2} \to 3^2P_{3/2}$   &&  198  &&  589   && 6.34  \\
$^{39}$K  &&  $4^2S_{1/2} \to 4^2P_{3/2}$   &&  358  &&  766   &&  1.81 \\
$^{85}$Rb &&  $5^2S_{1/2} \to 5^2P_{3/2}$   &&  770  &&  780   &&  1.63 \\
$^{133}$Cs&&  $6^2S_{1/2} \to 6^2P_{3/2}$   && 1270  &&  852.3 && 1.06 \\
$^{52}$Cr &&  $4^7S_{3} \to 4^7P_{4}$       &&  115  &&  425.6 &&  8.49 \\
$^{27}$Al &&  $3p^2P_{3/2} \to 3d^2D_{5/2}$ &&  85   &&  309.4 &&  57 \\
$^{69}$Ga &&  $4p^2P_{3/2} \to 4d^2D_{5/2}$ &&  382  &&  294.4 &&  127 \\
$^{115}$In && $5p^2P_{3/2} \to 5d^2D_{5/2}$ &&  634  &&  325.7 &&  78 \\
$^{107}$Ag && $5^2S_{1/2} \to 5^2P_{3/2}$   &&  601  &&  328 &&  76.8 \\
\hline\hline
\end{tabular}
\end{center} \caption{Dimmensionless atomic mass parameter corresponding
 to laser cooling transition for different elements.}
\end{table}

\begin{figure}[h]
\begin{center}
\includegraphics[width= 3.0 in]{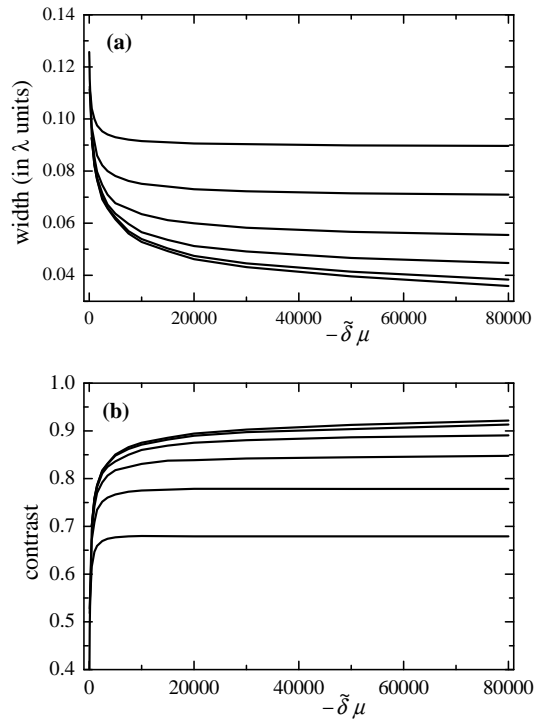}
\end{center}
\caption{\em Spatial FWHM width of localized atomic structures (a), and contrast (b) as function
 of parameter $\delta \, \mu$ for different field detuning ( $\delta/\gamma = -5, -10, -20, -40, -80, -160$ from top
 to bottom in (a) and from bottom to top in (b))
 in a model of atom with $j_g=1/2 \to j_e=3/2$ optical transition.} \label{fig2}
\end{figure}

\begin{figure}[h]
\begin{center}
\includegraphics[width= 3.0 in]{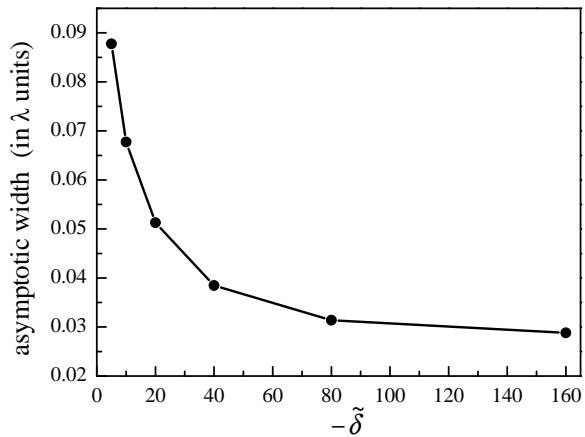}
\end{center}
\caption{\em Asymptotic FWHM width of localized atomic structures as function
 of light field detuning in a model of atom with $j_g=1/2 \to j_e=3/2$ optical transition.} \label{fig3}
\end{figure}

\begin{figure}[b]
\begin{center}
\includegraphics[width= 3.0 in]{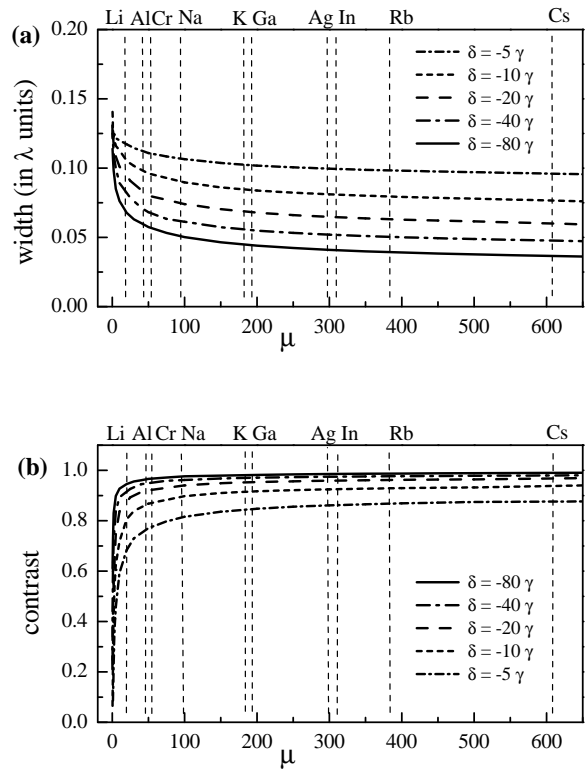}
\end{center}
\caption{\em Spatial FWHM width of localized atomic structures (a), and contrast (b) as function
 of parameter $\mu$ for different field detuning in a model of atom with $j_g=1 \to j_e=2$ optical transition.} \label{fig4}
\end{figure}

There are several physical parameters that are required to characterize a given laser
cooling situation. These are the atomic mass $M$, the wavelength $\lambda$, and the
natural line-width $\gamma$. In addition, there are two light-field parameters in the
low-field limit: detuning $\delta$, and saturation parameter $S$. It is possible to
choose reduced dimensionless units ($\hbar =1$, $k = 1$, $\gamma = 1$), so that the
dimensionless atomic mass $\tilde{M}$ can be defined from relation $\gamma/\omega_{R} =
2\tilde{M}$ \cite{Pru04}. This is so called a quasiclassical parameter that characterizes
an atom's kinetic in a light field. In particular, it describes the rate of kinetic
processes and evolution of atomic distribution function in momentum space, thus the
typical cooling time is order of $\tau =  \omega_{R}/(\gamma S)$.

We perform an analysis of the effects of atom localization in optical potential.
Contrast can be defined as the ratio of spatial distribution modulation depth to its
amplitude $C= h/H$ (Fig.\ref{fig1} (a)) as in \cite{Meschede03}. As it can be seen
directly from (\ref{wigner}), there are only two parameters that characterize the
stationary solution for an atomic density matrix. This is the detuning, that can be
measured in the units of natural line-width $\tilde{\delta} = \delta/\gamma$ and the
second one is non-dimensional parameter $\mu$:
\begin{equation}\label{mu}
  \mu = S \tilde{M} \, .
\end{equation}
This scale parameter makes it possible to implement the results of calculation for the
elements with allowed closed dipole optical transition having degeneration over the
angular momentum energy levels (table I).

Note, that in the secular approximation \cite{Dal91} the stationary solution is
characterized only by the ratio of optical potential depth to recoil energy
$U_0/\omega_R$ that is proportional to $\tilde{\delta} \mu$ in our notations. Thus we
first express the results in parameters $\tilde{\delta}$ and $\tilde{\delta} \mu$. As it
seen in the Fig. \ref{fig2} the differences between the curves become more significant
with the optical potential depth $U_0$ increasing, corresponding to out of framework
from secular approach. The width of the localized structures and the contrast tends to
an asymptotic constant value with increasing of the light field intensity, however that
depends on light field detuning.
 In order to find the asymptotic values for the width we fit curves with an
empirical low $w(x) = a + b/\sqrt{1 + c\, x}$. The results for asymptotic values $a$ are
shown in Fig. \ref{fig3}.

The Fig. \ref{fig4} represents the FWHM width and contrast of the spatial structures as
a function of parameter $\mu$ at different detunings for the atomic model with $j_g=1 \to
j_e=2$ optical transition. The FWHM width of the spatial structures is monotone decrease
for a large detuning and parameter $\mu$. In spite of the fact that the equilibrium
temperature is growing with the depth of optical potential, the localization of the atoms
becomes stronger with growth of these parameters. Additionally the contrast tends
towards its maximum value  Fig. \ref{fig4}(b). The dashed vertical lines on the
Fig.\ref{fig4} underline the limitations of the weak-field theory for different elements
from the table I in assumption $S < 0.5$. Note, that this is qualitative assumption. For
thorough analysis of the weak-field theory limits the solution of the quantum equation
for the total atomic density matrix is required with taking into account exited state
population. However, the width and contrast curves Fig.\ref{fig4} have a very strong
dependence on $\mu$, thus the localization effects remain rather significant for enough
small saturation parameter especially for "heavy" atoms table I.

\section{Conclusion}
We performed the fully quantum analysis of atomic localization obtained by laser cooling
in nonuniformly polarized low-intensity light field. Generally, conditions for a deep
laser cooling mismatch conditions for strong atomic localization that require deeper
optical potential and consequently leads to higher laser cooling temperature.
Additionally, in a deep optical potential the secular approximation (\ref{sec})
restricted by relation on light field detuning and the depth of optical potential. In
our treatment we had no such limitation allowing a more correctly describe the spatial
solution for atomic distribution function including localized and above-barrier motioned
atoms. The stationary solution of atoms is a function of the light field detuning
$\delta$ and the dimensionless parameter $\mu$ (\ref{mu}). We analyzed the width and
contrast of localized atomic structures as a function of these parameters. We showed
that the atomic structure width and contrast have a strong dependence on $\mu$ and tend
to constant values with an increasing optical potential dept that depends on the light
field detuning.

We demonstrated the applicability of laser cooling in far-off detuned deep optical
potential, created by a light field with polarization gradient, as a dissipative optical
mask for the purposes of atom lithography and nanofabrication for generation of spatially
localized atomic features with high contrast. This type of the light mask can be an
alternative method for creation of spatially localized atomic structures. The remarkable
distinction of this method from previous non-dissipative light mask is that the
suggested one is not sensitive to any aberration effects. More over, this type of
optical mask has no classical analog and can not be described by methods for classical
optics. The width and the contrast of localized atomic structures here is determined by
atomic energy dissipation mechanisms in the light field. Finally, we perform analysis of
the possible limits for the width and the contrast of localized atomic structures that
can in principle be reached by this type of the light mask.

\section{Acknowledgments}
This work was partially supported by RFBR (grants 05-02-17086, 04-02-16488, 05-08-01389)
and O.N.P. was supported by the grant MK-1438.2005.2.

\end{document}